\documentclass[10pt,a4paper,aps,prresearch,reprint,longbibliography,nobibnotes,floatfix,superscriptaddress,showpacs]{revtex4-2}

\usepackage[utf8]{inputenc}
\usepackage[english]{babel}

\usepackage{amsfonts}
\usepackage{amsmath}
\usepackage{amssymb}
\usepackage[dvipsnames,svgnames,table]{xcolor}
\usepackage{graphicx}
\usepackage{fancyhdr}
\usepackage{float}
\usepackage{braket}
\usepackage{float}
\usepackage[caption=false]{subfig}
\usepackage{tikz}  
\usepackage{chngcntr}
\usepackage{ulem}

\usepackage{mathrsfs}

\usepackage[dvipsnames]{xcolor}

\usepackage{hyperref}
\usepackage{placeins}
\hypersetup{
    pdfstartview={FitH},
    colorlinks=true,    
    linkcolor=NavyBlue, 
    citecolor=Maroon,   
    filecolor=NavyBlue, 
    urlcolor=NavyBlue   
}

\makeatletter
\def\footnoterule{\kern -10pt
    \hrule \@width 100pt \kern 10pt} 
\makeatother

\usepackage{tipa}

\begin{document}

\title{Nonlinear Interaction and Topological Selection Rules in two Skyrmion Merging: Emergence of Skyrmionium Textures}

\author{M. Al Busaidi}
\affiliation{Department of Physics, College of Science, Sultan Qaboos University, P.O. Box 36, PC 123, Muscat, Oman}

\author{T. Hakioglu}
\affiliation{Energy Institute and Department of Physics, Istanbul Technical University, Maslak 34469, Istanbul, Türkiye}

\author{R. Sbiaa}
\email{rachid.sbiaa@squ.edu.om}
\affiliation{Department of Physics, College of Science, Sultan Qaboos University, P.O. Box 36, PC 123, Muscat, Oman}

\date{\today}

\begin{abstract}
The nonlinear interaction between magnetic skyrmions at short separation remains weakly understood despite its promise in high-tech applications in topological spin manipulations. In particular, the physical conditions governing whether two interacting skyrmions merge into a conventional skyrmion or evolve into a stable skyrmionium are still unclear.
Here, we investigate as an example, the merging dynamics of two Néel skyrmions within a micromagnetic framework by combining analytical modelling with simulations based on Landau–Lifshitz–Gilbert dynamics. We systematically examine the role of exchange, Dzyaloshinskii–Moriya interaction, anisotropy, and magnetostatic energies during the nonlinear relaxation process while varying the initial separation, polarity, winding number, and the skyrmion size.
We demonstrate computationally that the final magnetic state is governed by a topological selection rule determined by the combined polarity–winding number configuration of the interacting skyrmions. Among all possible combinations, a stable skyrmionium emerges exclusively for skyrmions with identical polarity and opposite winding number at intermediate overlap, whereas all remaining configurations relax into a single skyrmion. Furthermore, we identify a critical skyrmion radius below which the composite state becomes unstable and collapses, revealing that skyrmionium formation requires not only the appropriate topology but also a minimum geometrical size.
Analysis of the individual energy contributions shows that the transition is driven by the cooperative relaxation of exchange frustration together with the redistribution of Dzyaloshinskii–Moriya and magnetostatic energies, leading to the spontaneous formation of a double-domain-wall configuration. The resulting skyrmionium exhibits persistent breathing oscillations about a well-defined equilibrium radius, demonstrating that it represents a dynamically stable metastable state rather than a transient intermediate.
These promising results establish a predictive physical framework for controlling nonlinear skyrmion merging and provide practical design rules for engineering skyrmionium textures in chiral magnetic nanostructures.
\end{abstract}

\maketitle

\section{Introduction}
\label{sec:introduction}

Magnetic skyrmions are nanoscale, spatially localized spin textures characterized
by a nontrivial mapping of the magnetization field onto the unit sphere
\cite{Bogdanov1989,Muhlbauer2009,Yu2010,Nagaosa2013,Fert2017}.
Their particle-like character, nanoscale dimensions, and ability to be created,
annihilated, and transported by external stimuli have established them as
promising information carriers for high-density magnetic memory, logic, and
neuromorphic architectures
\cite{Fert2013,Sampaio2013,Fert2017,EverschorSitte2018,kumar2024emulation}.
In ultrathin ferromagnetic films with broken inversion symmetry, their
stabilization is governed by the competition between exchange interaction,
perpendicular magnetic anisotropy, interfacial DMI, magnetostatic coupling, and external magnetic fields
\cite{Bogdanov1994,Bogdanov1999,Rohart2013,Woo2016,Buettner2018,Ding2019,Caretta2020,naabi2022skyrmion,Kuepferling2023}.
The resulting balance supports localized chiral textures over a finite region
of the material-parameter space and simultaneously controls their size,
internal structure, and dynamical response.

Beyond the properties of an isolated skyrmion, interactions between multiple
topological textures are central to collective skyrmion dynamics and to the
operation of densely packed skyrmionic devices \cite{lin2013,Loudon2018,Capic2019,Reichhardt2022,Warda2023dynamics}. At sufficiently large separations, skyrmions can often be described approximately as interacting
quasiparticles, with forces originating from the overlap of their exponentially
decaying magnetic tails, dipolar fields, boundary-induced distortions, and
material inhomogeneities
\cite{Reichhardt2022}.
This weak-overlap description becomes inadequate when the center-to-center
separation approaches the skyrmion diameter. In this strongly nonlinear regime,
the domain walls and core regions overlap, the individual profiles lose their
identity, and large magnetization gradients develop within the interaction
region. The resulting evolution can no longer be represented by a simple
pairwise potential between rigid particles. Instead, the complete magnetic
texture undergoes substantial deformation, chiral reconstruction, and
redistribution of exchange, DMI, anisotropy, and magnetostatic energies.
Depending on the initial configuration and the accessible relaxation pathway,
closely interacting skyrmions may remain separated, merge into a single
skyrmion, form an extended composite texture, or ultimately collapse
\cite{Zhang2015,Koshibae2016,Desplat2017,Loudon2018,Yang2023}.

A particularly important composite state is the skyrmionium, also referred to
as a $2\pi$ skyrmion or target-like skyrmion. It consists of two concentric
domain walls separating a central core, an intermediate annular domain, and
the surrounding ferromagnetic background
\cite{Zhang2016,Zheng2017,Zhang2018}.
Unlike a conventional skyrmion, whose magnetization typically reverses once
between the center and the background, the radial magnetization profile of a
skyrmionium undergoes two successive reversals. The two constituent windings
carry compensating topological contributions, giving the ideal skyrmionium a
vanishing net topological charge. Consequently, its gyrotropic response can be
strongly suppressed, allowing nearly longitudinal motion and potentially
higher propagation velocities than those of ordinary ferromagnetic skyrmions
\cite{Kolesnikov2018,Gobel2019}. These characteristics have motivated proposals
for using skyrmioniums as information carriers in racetrack devices and
multistate spintronic platforms
\cite{Zhang2016,Gobel2019,Yang2023}.

Several mechanisms for creating skyrmionium textures have been proposed,
including local spin-current injection, magnetic-field pulses, geometric or
curvature engineering, optical excitation, and controlled conversion from
pre-existing skyrmions
\cite{Zhang2016,Pylypovskyi2018,Gobel2019,Yang2023}.
Experimental observations have further demonstrated that skyrmionium-like
textures can be stabilized in confined chiral nanostructures and magnetic
heterostructures
\cite{Zheng2017,Zhang2018}.
Most previous investigations, however, have focused on the generation,
current-driven transport, or field-induced transformation of an already formed
skyrmionium. Comparatively less attention has been devoted to the spontaneous
formation of a skyrmionium through the direct nonlinear merging of two initially
distinct skyrmions. In particular, the combined roles of initial separation,
polarity, winding number, and skyrmion size in selecting the final magnetic state
have not yet been established within a unified dynamical framework.

This distinction is important because the topological charge alone does not
fully specify the local compatibility of two merging textures. For an axially
symmetric skyrmion, the topological charge is determined by the product of the
core polarity and the winding number, whereas the detailed connection
of the spins across an overlap region also depends on how these quantities are
distributed between the two interacting objects. Two configurations having
related global topological charges may therefore follow qualitatively different
relaxation pathways when their domain walls overlap. In one case, the
in-plane magnetization may connect continuously and reconstruct into a
double-domain-wall state; in another, incompatible winding may generate large
exchange frustration and favour relaxation into a single-core structure.
Accordingly, a predictive description of skyrmion merging must account not only
for the total topological charge, but also for the separate polarity and
winding number of each interacting texture.

A second unresolved issue concerns the geometrical stability of the merged
state. Even when the initial topology permits a ring-like reconstruction, the
formation of a stable skyrmionium requires sufficient space to sustain both an
inner and an outer domain wall. If the initial skyrmions are too small, the two
walls may become strongly compressed, causing the annular domain to disappear
and the composite state to transform into a conventional skyrmion or collapse.
The existence of a minimum size for merger-induced skyrmionium formation is
therefore expected from energetic considerations, but its relationship to the
nonlinear relaxation process and to the individual micromagnetic energy
contributions requires systematic analysis.

In this work, we investigate the nonlinear interaction and merging dynamics of
two N\'eel-type skyrmions in an ultrathin ferromagnetic film. We combine an
analytical description of the equilibrium skyrmion length scale with
time-dependent micromagnetic simulations based on the
Landau--Lifshitz--Gilbert equation. The initial center-to-center separation,
polarity, winding number, and skyrmion radius are varied systematically in order to
identify the physical variables that select the final relaxed state. Rather
than treating the skyrmions as rigid quasiparticles, we resolve the complete
magnetization dynamics and follow the time evolution of the exchange,
anisotropy, DMI, and demagnetizing-energy contributions during the merging
process.

Our results reveal a topological selection rule for merger-induced skyrmionium
formation. Specifically, for $p_i=\pm1$ and $n_i=\pm1$ denoting the polarity and
winding number of the $i$th skyrmion, respectively, the observed selection
rule can be expressed as $p_1p_2=+1$ and $n_1n_2=-1$, corresponding to
identical polarity and opposite winding number. For the investigated N\'eel
configurations, a stable skyrmionium is obtained when two partially overlapping
skyrmions satisfy this condition, allowing their in-plane magnetization to
reconstruct continuously into a double-domain-wall texture. Other polarity--winding number
combinations relax predominantly toward a single-skyrmion state. We further
identify a critical initial skyrmion size below which the ring-like composite
configuration cannot be maintained and ultimately collapses. The decomposition
of the total energy shows that skyrmionium formation is driven primarily by
the removal of the exchange frustration in the initial overlap region, while DMI,
anisotropy, and magnetostatic interactions reorganize cooperatively to stabilize
the final concentric structure. In the low-damping regime, the relaxed
skyrmionium exhibits a bounded radial breathing mode around a well-defined
equilibrium radius, providing an additional dynamical signature of its
metastability.

These findings establish a unified physical picture of short-range skyrmion
merging in which topology, geometry, and energy relaxation jointly determine
the final magnetic configuration. The resulting selection criteria provide
practical guidelines for engineering skyrmionium textures through controlled
skyrmion interactions and may support the development of reconfigurable
topological spintronic devices.

\section{Methodology}

The magnetic configurations considered in this work are described within the micromagnetic framework~\cite{Fert2013,Nagaosa2013}. In this formulation, the total energy of the system is given by a micromagnetic functional

\begin{equation}
\begin{aligned}
E[\mathbf{m}] = & \int d^2 r \Big[ A |\nabla \mathbf{m}|^2 + K_{\mathrm{eff}} (1 - m_z^2) \\
& + D \big( m_z \nabla \cdot \mathbf{m}_{\perp} - \mathbf{m}_{\perp} \cdot \nabla m_z \big) + E_Z + E_{\mathrm{dem}} \Big],
\end{aligned}
\label{eq:energy}
\end{equation}

where $\mathbf{m}(\mathbf{r})=(m_x,m_y,m_z)$ is the spatially varying unit magnetization vector field satisfying $|\mathbf{m}(\mathbf{r})|=1$, $A>0$ is the exchange stiffness constant that energetically favors smooth spatial variations of the magnetization, $K_{\mathrm{eff}}>0$ is the effective perpendicular magnetic anisotropy constant favoring out-of-plane spin alignment ($m_z=\pm1$), and $D$ is the interfacial Dzyaloshinskii--Moriya interaction constant responsible for stabilizing chiral N\'eel-type spin textures. Here $\mathbf{m}_{\perp}$ represents the in-plane component of the magnetization. The last two terms correspond to the Zeeman energy and magnetostatic (demagnetizing) energy, respectively~\cite{Bogdanov1989,Bogdanov1994}.

The first term,
$A|\nabla \mathbf{m}(\mathbf{r})|^2$,
represents the exchange interaction, where the positive exchange stiffness constant $A>0$ penalizes rapid spatial variations of the magnetization field. Consequently, this contribution favors a smooth and nearly collinear alignment of neighboring spins and opposes the strong twisting required for skyrmion formation. Nevertheless, the exchange interaction is essential for stabilizing a continuous and regular spin texture.

The second term,
$K_{\mathrm{eff}}[1-m_z^2(\mathbf{r})]$,
corresponds to the effective perpendicular magnetic anisotropy. For positive $K_{\mathrm{eff}}>0$, the energy is minimized when $m_z=\pm1$, thereby favoring out-of-plane magnetization. This contribution stabilizes the ferromagnetic background while allowing the existence of a localized reversed core at the skyrmion center.

The third term, proportional to the Dzyaloshinskii--Moriya interaction (DMI) constant $D$, induces chiral spin textures by energetically favoring a fixed rotational sense of the magnetization. This interaction promotes twisted configurations such as N\'eel or Bloch skyrmions and counteracts the tendency of the exchange interaction to maintain a spatially uniform magnetization state.

The Zeeman energy,
\begin{equation}
E_Z = -\mu_0 M_s \mathbf{H}\cdot\mathbf{m}(\mathbf{r}),
\end{equation}
describes the coupling between the magnetization and an externally applied magnetic field $\mathbf{H}$. This term tends to align the spins along the field direction and can significantly influence the skyrmion size and stability.

Finally, the demagnetizing energy,
$E_{\mathrm{dem}}$,
originates from long-range dipolar interactions and plays an important role in thin magnetic films by favoring flux-closure configurations. This contribution affects the equilibrium shape and stabilization of skyrmionic textures.

Overall, skyrmion formation results from the competition among these energy contributions: the exchange interaction favors spatially smooth magnetization, the anisotropy stabilizes out-of-plane alignment, the DMI induces chiral twisting, while the Zeeman and dipolar interactions further control the size, stability, and equilibrium configuration of the skyrmion state~\cite{Bogdanov1994,Nagaosa2013}.

The time evolution of the magnetization is governed by the Landau--Lifshitz--Gilbert (LLG) equation\cite{LLG-1}, 
\begin{equation}
\frac{\partial \mathbf{m}}{\partial t}
=
-\gamma \,
\mathbf{m}
\times
\mathbf{H}_{\mathrm{eff}}
+
\alpha \,
\mathbf{m}
\times
\frac{\partial \mathbf{m}}{\partial t},
\label{eq:LLG}
\end{equation}
where $\gamma$ is the gyromagnetic ratio and $\alpha$ is the Gilbert damping parameter normalized by the saturation magnetization. 

The first term on the right-hand side of Eq.~(\ref{eq:LLG}) describes the precessional motion of the magnetization around the effective magnetic field $\mathbf{H}_{\mathrm{eff}}$, while the second term accounts for damping and drives the system toward a lower-energy equilibrium state. In the absence of damping, the magnetization undergoes perpetual precession without relaxation. The damping contribution is therefore essential for energy dissipation and for the stabilization of magnetic textures such as skyrmions and skyrmionium states.

The effective magnetic field is obtained from the variational derivative of the total energy functional as,

\begin{equation}
\mathbf{H}_{\mathrm{eff}}
=
-\frac{1}{\mu_0 M_s}\frac{\delta E}{\delta \mathbf{m}},
\label{eq:heff}
\end{equation}
where $\mu_0$ is the vacuum permeability. 

Since the magnetization vector is normalized to unity $|\mathbf{m}|=1$, the LLG equation governs only the change in its orientation and not in its magnitude. As a result, the dynamical evolution described by Eq.~(3) corresponds to a combination of field-driven precession and damping-induced relaxation. This framework is fundamental for describing the nucleation, propagation, deformation, interaction, and relaxation of the topological magnetic textures, including the formation and stabilization of skyrmions and skyrmionium states.To describe an isolated skyrmion structure, we employ an axially symmetric ansatz in polar coordinates $(r,\phi)$:

\begin{equation}
\mathbf{m}(r,\phi) = 
\begin{pmatrix}
\sin f(r)\cos(n\phi+\chi) \\
\sin f(r)\sin(n\phi+\chi) \\
\cos f(r)
\end{pmatrix}
\label{eq:ansatz}
\end{equation}

where $f(r)$ is the radial profile function, $n$ denotes the winding number, and $\chi$ represents the helicity. The function $f(r)$ describes how the magnetization varies radially from the skyrmion center to the ferromagnetic background. In particular, the out-of-plane component is given by $m_z=\cos f(r)$, so that the value of $f(r)$ at the origin determines the core orientation, while its asymptotic value fixes the background magnetization. Thus, the ansatz provides a compact description of a localized twisted magnetic texture with axial symmetry.

The winding number index $n$ specifies how many times the in-plane magnetization winds as the azimuthal angle $\phi$ changes from $0$ to $2\pi$. The helicity $\chi$ fixes the orientation of the in-plane magnetization relative to the radial direction. In particular, $\chi=0$ or $\pi$ corresponds to N\'eel-type skyrmions, for which the in-plane magnetization points radially inward or outward, while $\chi=\pm\pi/2$ corresponds to Bloch-type skyrmions, for which the in-plane component is tangential to circles centered on the core~\cite{Rohart2013}. Therefore, the helicity distinguishes different internal spin structures without changing the topological class of the texture.

The topology of the magnetic texture is characterized by the topological charge $Q$ [1] which measures how many times the magnetization wraps the unit sphere in the two dimensional  $r, \phi$ plane. For a skyrmion, the local magnetization explores all directions from the north pole to the south pole of the unit sphere, producing a nonzero integer topological charge. This quantity is a topological invariant and remains unchanged under smooth deformations of the texture.

For axially symmetric configurations, the topological charge is determined by
the polarity and the winding number. The polarity $p$ describes the orientation
of the skyrmion core and is defined by the sign of the out-of-plane
magnetization at the center, namely $p=+1$ for an upward core and $p=-1$
for a downward core. For the axially symmetric ansatz in
Eq.~(\ref{eq:ansatz}), the topological charge can be written as $Q=\frac{n}{2}\left[\cos f(0)-\cos f(\infty)\right]
=\frac{n}{2}\left[m_z(0)-m_z(\infty)\right].$
Therefore, the sign of $Q$ is determined by the winding number together with
the relative orientation of the core and the ferromagnetic background. Hence,
reversing either the polarity or the winding number reverses the sign of $Q$,
whereas changing the helicity alone does not modify the topological charge.
Chirality describes the handedness of the spin rotation in a skyrmion and is
determined by the sign of the DMI, which selects the energetically preferred
twisting direction. While the helicity $\chi$, as defined in
Eq.~(\ref{eq:ansatz}), specifies the in-plane spin orientation, chirality
reflects the material-imposed rotational sense. Together, the skyrmion
structure is characterized by four quantities: polarity (core orientation),
winding number, helicity (spin orientation), and chirality (DMI-driven
handedness).

\section{Result and Discussion}
Based on these observations, we numerically run micromagnetic simulations of
Eq.~(\ref{eq:LLG}) for a single isolated skyrmion $\mathbf{m}(\mathbf{r})$,
as well as for two interacting skyrmions separated by a displacement
$\mathbf{\Delta}$. Each individual skyrmion texture satisfies
$|\mathbf{m}_1(\mathbf{r})|=|\mathbf{m}_2(\mathbf{r})|=1$. The combined
initial magnetization texture is constructed from the two skyrmion profiles
and normalized pointwise according to
$\mathbf{M}(\mathbf{r})
=
\frac{\mathbf{m}_1(\mathbf{r})+\mathbf{m}_2(\mathbf{r}+\mathbf{\Delta})}
{\left|\mathbf{m}_1(\mathbf{r})+\mathbf{m}_2(\mathbf{r}+\mathbf{\Delta})\right|},$
such that $|\mathbf{M}(\mathbf{r})|=1$ throughout the magnetic film. This
normalized configuration is then used as the initial state for the subsequent
LLG relaxation.\\
The micromagnetic simulations were performed using material parameters representative of Co/Pt ultrathin ferromagnetic films~\cite{MoreauLuchaire2016,Woo2016,Rohart2013} to allow us a direct comparison with experiments. The saturation magnetization is set to $M_s = 6 \times 10^5~\mathrm{A/m}$. Furthermore, the parameters of the Eq.(\ref{eq:energy}) are fixed as, the exchange stiffness $A = 7 \times 10^{-12}~\mathrm{J/m}$, the interfacial Dzyaloshinskii--Moriya interaction strength $D = 2 \times 10^{-3}~\mathrm{J/m^2}$, and the uniaxial anisotropy constant $K_u = 6.5 \times 10^5~\mathrm{J/m^3}$. The corresponding effective perpendicular anisotropy is given by$ K_{\mathrm{eff}}=K_u-\frac{1}{2}\mu_0M_s^2,$ which accounts for the reduction caused by magnetostatic demagnetizing effects. The anisotropy axis was  initially chosen as $\mathbf{u}=(0,0.01,1)$, corresponding to an almost perfectly out-of-plane easy axis  frustrated along the $y$-direction by a small amount.

The simulations presented in this section are designed to test the physical hypothesis proposed in Sec.I, namely that the final state produced by nonlinear skyrmion merging is governed by two complementary mechanisms. First, the polarity--winding number combination determines the topologically allowed relaxation pathway. Second, even for a topologically favourable configuration, the formation of a stable skyrmionium requires a sufficiently large initial skyrmion size. The following results are therefore organized to examine these two mechanisms separately and to establish their combined role in selecting the final magnetic configuration.

To test this hypothesis, we first consider the interaction between two identical Néel skyrmions of equal radius $\vert \mathbf{R} \vert$, while systematically varying their center-to-center separation $\mathbf{\Delta}$ within the short-distance regime.

We define the static interaction between two skyrmions based on the energy
functional in Eq.~(\ref{eq:energy}). The interaction energy is defined as
\begin{equation}
U(\mathbf{\Delta})
=
E[\mathbf{M}]
-
\left(
E[\mathbf{m}_1]+E[\mathbf{m}_2]
\right),
\label{interaction}
\end{equation}
where $\mathbf{\Delta}$ denotes the displacement vector between the two
skyrmions. In the evaluation of $E[\mathbf{M}]$, the energy is calculated
from the LLG solution initialized with the combined magnetization texture
$\mathbf{M}(\mathbf{\Delta})$.

The simulations are performed in the absence of external magnetic fields and spin-polarized currents so that the observed dynamics originates solely from the intrinsic competition between exchange, Dzyaloshinskii--Moriya interaction, perpendicular magnetic anisotropy, and magnetostatic energies. The analysis is subsequently extended to different polarity and winding number combinations in order to determine how topology modifies the relaxation pathway and the resulting equilibrium state.

Before investigating the interaction dynamics, it is necessary to identify the physically relevant equilibrium skyrmion radius for the chosen material. Rather than selecting the initial radius arbitrarily, it is determined from the intrinsic material parameters through its dependence on the DMI strength. This choice is motivated by the fact that the DMI is the primary parameter controlling the equilibrium skyrmion size because it directly stabilizes the chiral rotation of the magnetization and determines the proximity of the system to the instability threshold~\cite{Rohart2013}. Although the exchange stiffness $A$ and the effective anisotropy $K_{\mathrm{eff}}$ define the characteristic magnetic length $\Delta=\sqrt{\frac{A}{K_{\mathrm{eff}}}}$~\cite{Rohart2013}, these quantities remain fixed for a given material and therefore do not describe the evolution of the equilibrium skyrmion size. In contrast, the skyrmion radius increases rapidly as the DMI approaches the critical value $D_c=\frac{4}{\pi}\sqrt{A K_{\mathrm{eff}}}$~\cite{Rohart2013}, reflecting the progressive reduction of the effective domain-wall energy. Consequently, determining the initial radius from the equilibrium $R(D)$ relation ensures that the simulated skyrmions correspond to physically stable configurations and that the subsequent interaction dynamics is governed by the intrinsic balance between exchange, anisotropy, DMI, and magnetostatic interactions rather than by an arbitrary geometrical choice.

\begin{figure}[t!]
    \centering
    \includegraphics[width=0.9\linewidth]{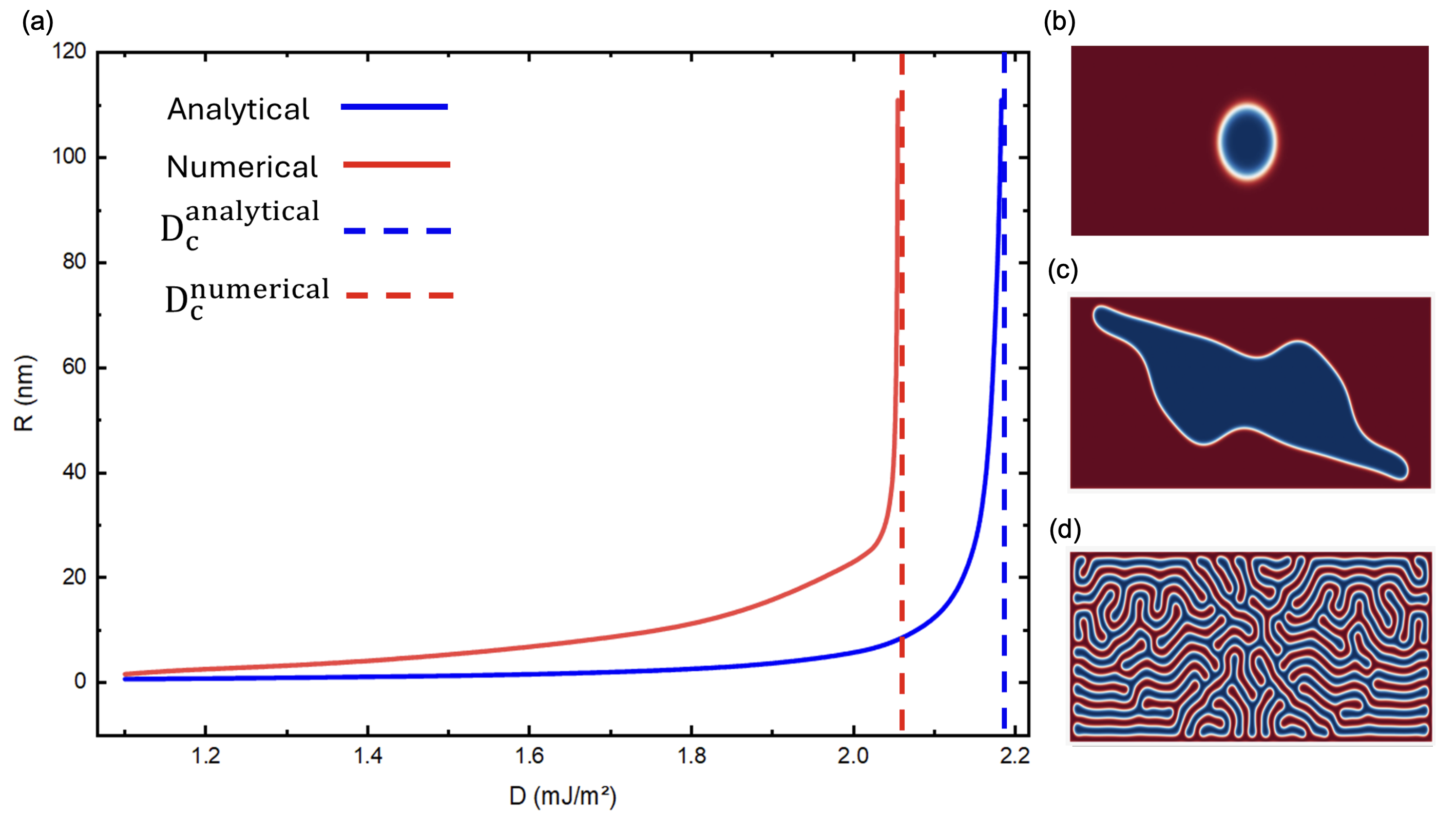}
\caption{
Dependence of the skyrmion radius and magnetic configuration on the
Dzyaloshinskii--Moriya interaction (DMI) strength $D$.
(a) Skyrmion radius $R$ as a function of $D$, comparing the analytical model
(blue curve) with the micromagnetic simulations (red curve). The vertical
dashed lines indicate the analytical and numerical critical DMI strengths,
$D_c^{\mathrm{analytical}}\approx2.19~\mathrm{mJ/m^2}$ and
$D_c^{\mathrm{numerical}}\approx2.055~\mathrm{mJ/m^2}$, respectively.
(b) Magnetization configuration below the numerical critical DMI, showing a
stable isolated and approximately circular skyrmion.
(c) Magnetization configuration near the numerical critical DMI, where the
isolated skyrmion expands and becomes strongly distorted.
(d) Magnetization configuration above the numerical critical DMI, showing
the transition from an isolated skyrmion to an extended multidomain state.
}
\end{figure}
Figure~1 establishes the physically relevant single skyrmion
radius used throughout the present study by comparing the analytical
circular-domain-wall model with full micromagnetic simulations. As shown in
Fig.~1(a), both approaches predict a monotonic expansion
of the skyrmion with increasing DMI strength
$D$. At comparatively low DMI values, approximately
$D\lesssim 1.8~\mathrm{mJ\,m^{-2}}$, the analytical and numerical results
remain in reasonable qualitative agreement. In this regime, the skyrmion is
compact relative to the lateral dimensions of the sample, its profile remains
approximately axisymmetric, and the circular-domain-wall approximation
captures the dominant competition among exchange, perpendicular anisotropy,
and DMI.

Within the analytical model, the equilibrium skyrmion radius is described by
\begin{equation}
\begin{aligned}
R(D) &= \frac{\ell_{\rm DW}}{\sqrt{2\left(1-D/D_{c}^{\mathrm{an}}\right)}}, \\[6pt]
\ell_{\rm DW} &= \sqrt{\frac{A}{K_{\mathrm{eff}}}}, \qquad D_{c}^{\mathrm{an}} = \frac{4}{\pi}\sqrt{A K_{\mathrm{eff}}}.
\end{aligned}
\end{equation}

where $\ell_{\rm DW}$ is the characteristic magnetic length and
$D_{c}^{\mathrm{an}}$ is the critical DMI strength predicted by the
analytical circular-domain-wall model. The corresponding effective
domain-wall energy is~\cite{Rohart2013}
\begin{equation}
\sigma_{\mathrm{DW}}
=
4\sqrt{A K_{\mathrm{eff}}}-\pi D.
\end{equation}
Accordingly, as $D$ approaches $D_{c}^{\mathrm{an}}$ from below,
$\sigma_{\mathrm{DW}}$ decreases toward zero and the analytical model
predicts a rapid increase of the skyrmion radius. For the material parameters
considered here, this gives
$D_{c}^{\mathrm{an}}\simeq2.19~\mathrm{mJ\,m^{-2}}$. The divergence of
$R(D)$ at this value is a consequence of the analytical circular-domain-wall
approximation and should not be interpreted as a physically observable
infinitely large isolated skyrmion~\cite{Rohart2013}.

The micromagnetic simulations exhibit the same qualitative expansion with
increasing $D$, but yield a lower numerical threshold,
$D_{c}^{\mathrm{num}}\simeq2.055~\mathrm{mJ\,m^{-2}}$. The difference between
$D_{c}^{\mathrm{an}}$ and $D_{c}^{\mathrm{num}}$ should not be attributed
solely to the finite lateral dimensions of the simulated sample. The
analytical expression is obtained from an idealized circular-domain-wall
description, whereas the micromagnetic simulations allow the complete
magnetization texture to relax and include the demagnetizing field, finite
geometry, edge-induced distortions, and nonlinear changes of the skyrmion
profile~\cite{Vansteenkiste2014,GarciaSanchez2016}. 

The spin configurations shown in Figs.~1(b)--1(d) illustrate this evolution.
Below $D_{c}^{\mathrm{num}}$, Fig.~1(b), the magnetization forms a localized
and approximately axisymmetric isolated skyrmion. Near the numerical
threshold, Fig.~1(c), the texture expands and becomes strongly distorted,
indicating that the circular isolated-skyrmion description is no longer
appropriate. For the larger DMI value considered in Fig.~1(d), the simulation
evolves from the isolated skyrmion into an extended modulated magnetic state.
We emphasize that Fig.~1(d) represents the state obtained for the specific
material parameters and computational geometry considered here and is not
intended as a general statement about the magnetic ground state for all
$D>D_c$.

For the two-skyrmion calculations presented below, the initial skyrmion
radius is therefore extracted from the stable numerical branch. Using the
material parameters specified above and
$D=2.00~\mathrm{mJ\,m^{-2}}$, corresponding to
$D/D_{c}^{\mathrm{num}}\simeq0.973$, the relaxed micromagnetic configuration
remains an isolated skyrmion with radius $R\simeq23~\mathrm{nm}$. This value
is adopted as the reference skyrmion radius in the subsequent interaction
analysis because it represents a stable relaxed configuration below the
numerical instability threshold.

An accurate determination of $R$ is also essential for distinguishing the
different interaction regimes between two skyrmions. The degree of overlap
is characterized by the normalized center-to-center separation $\Delta/R$. In the
strong-overlap regime, the two initial textures undergo substantial
reconstruction and may form either a skyrmionium or a single skyrmion,
depending on their individual polarity and winding-number parameter set. At intermediate
separations, the initially distinct skyrmions interact dynamically and evolve
toward the final relaxed configuration without immediately forming a merged
texture. At sufficiently large normalized separations, $\Delta/R\gtrsim2$, the
initial overlap is weak and the interaction occurs over a longer dynamical
timescale.
Figure~1 therefore serves two purposes. First, the comparison between the analytical and numerical results establishes the range in which the circular-domain-wall model provides a useful leading-order description of the skyrmion size. The spin textures in Figs.~1(b)--1(d) provide a direct visual assessment of the progressive loss of the approximately axisymmetric isolated skyrmion configuration near the numerical threshold. Second, Fig.~1 identifies the stable numerical branch from which the reference radius used in the subsequent two-skyrmion interaction analysis is extracted. The analytical model therefore provides the relevant physical scaling, whereas the micromagnetic simulations determine the relaxed skyrmion size and configuration for the complete system~\cite{Rohart2013,Leonov2015,Vansteenkiste2014}.
\begin{figure}[htbp]
    \centering
    \includegraphics[width=1\linewidth]{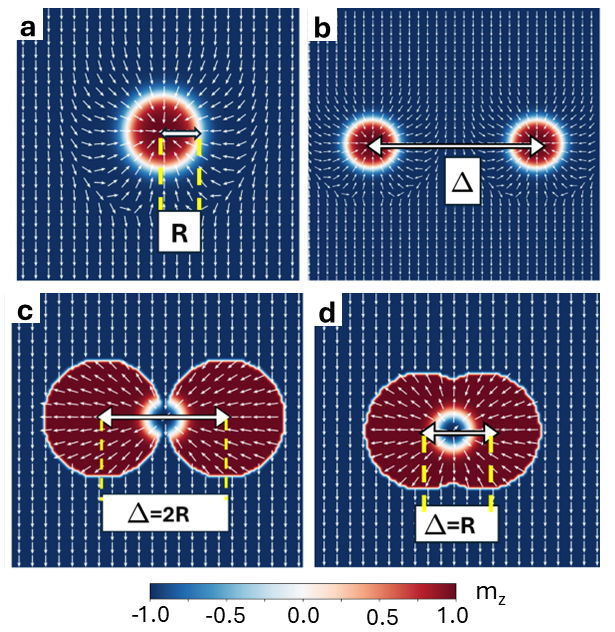}
\caption{Schematic illustration of the definition of the skyrmion radius $R$ and the center-to-center separation $\Delta$ between two N\'eel skyrmions, together with their evolution at different interaction distances. 
(a) Single skyrmion configuration used to define the radius $R$, measured from the center to the domain wall where $m_z = 0$. 
(b) Two well-separated skyrmions with center-to-center distance $\Delta$, where the interaction is negligible. 
(c) Intermediate regime at $\Delta= 2R$, where the skyrmions begin to interact while maintaining their individual identities. 
(d) Strong interaction regime at $\Delta = R$, where the skyrmions significantly partially overlap.
The color map represents the out-of-plane magnetization component $m_z$, while arrows indicate the in-plane spin configuration.
}
\end{figure}

Figure~2(a) illustrates the radius $R$ of an isolated N\'eel skyrmion, measured from the skyrmion center to the domain wall. Figure~2(b) then illustrates two well-separated skyrmions with a
center-to-center separation $\Delta$, for which the overlap between their spin textures is weak. As the separation decreases, specifically near $\Delta \simeq 2R$, as shown in Fig.~2(c), the outer regions of the two spin textures begin to overlap, marking the transition from the weak-interaction regime to a regime in which short-range contributions become increasingly important. In this regime, exchange and DMI play central roles in the ensuing reconstruction. The initial overlap produces a strongly distorted magnetization field with large local gradients. During relaxation, exchange favors the removal of these sharp gradients, whereas DMI promotes a continuous rotation of the magnetization with the preferred chirality. Their interplay therefore
facilitates the reconstruction of the overlapping textures into a smoother magnetic configuration.

This effective attraction drives the skyrmions toward each other and, for all investigated combinations of polarity and azimuthal winding number at $\Delta \simeq 2R$, ultimately leads to merging into a single-skyrmion state rather than the formation of a stable composite skyrmionium. The specific case $Q_1=Q_2=1$ provides a particularly instructive example: although the initial total topological charge is $Q=Q_1+Q_2=2$, the collision and subsequent reconstruction yield a single skyrmion with $Q_f=1$.

For a smooth magnetization field, the topological charge
$
Q=\frac{1}{4\pi}\int
\mathbf{m}\cdot
\left(
\partial_x\mathbf{m}\times\partial_y\mathbf{m}
\right)\,d^2r
$
is invariant under continuous deformations that preserve the relevant boundary conditions.
Therefore, the observed change from $Q=Q_1+Q_2=2$ to $Q_f=1$ cannot be attributed simply
to smooth dissipative relaxation. Instead, it indicates a topological transition occurring
during the strongly nonlinear merging process, where large and highly localized magnetization
gradients develop in the interaction region. In the discretized micromagnetic system, such a
localized reconstruction provides a pathway by which the initial topological configuration can
change and subsequently relax into the lower-energy single-skyrmion state \cite{lin2013, Nagaosa2013}.

\begin{figure}[bt!]
    \centering
    \includegraphics[width=1\linewidth]{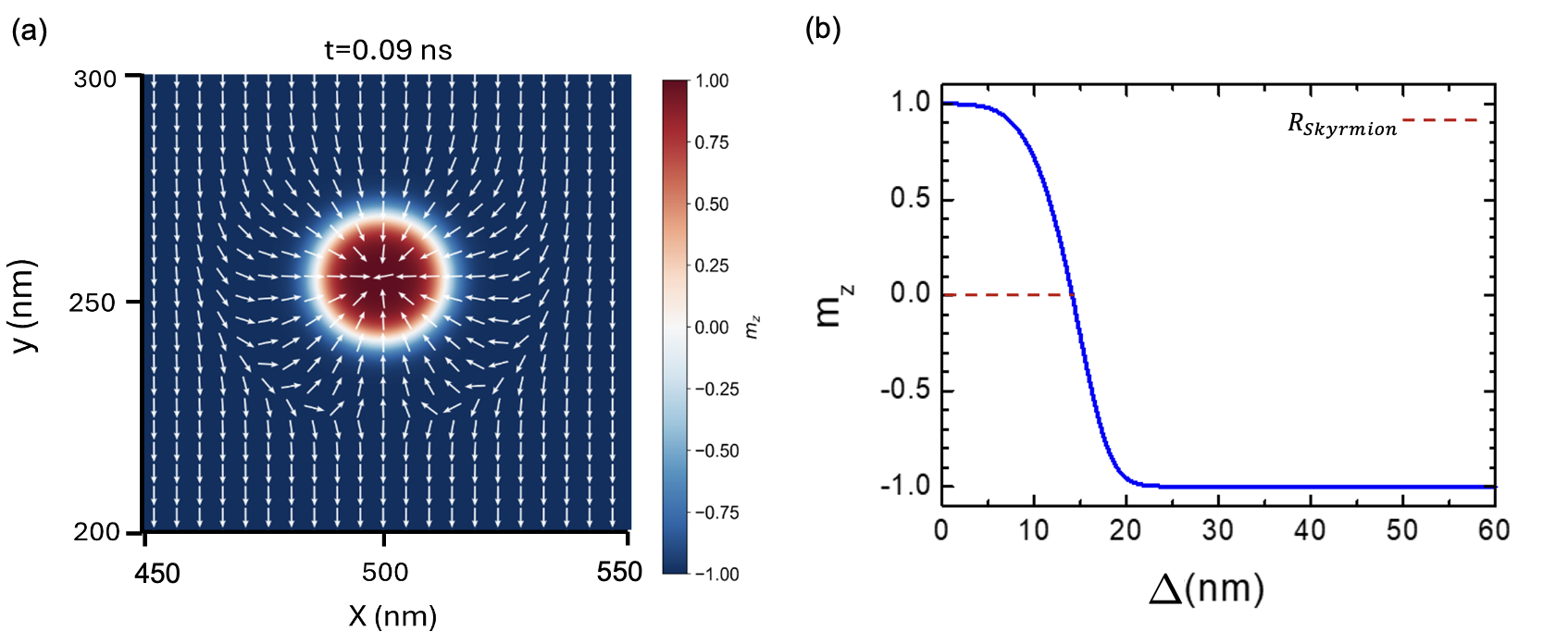}
\caption{
Formation of a  skyrmion resulting from the merging of two initial N\'eel skyrmions of radius $R \approx 23$ nm. 
(a) Snapshot of the magnetization configuration at $t = 0.09$ ns after merging. The two initial skyrmions have coalesced into a single skyrmion with a preserved chiral structure.
(b) Radial profile $m_z(r)$ extracted through the center of the merged structure.}
\end{figure}

\begin{figure}[bt!]
    \centering
    \includegraphics[width=1\linewidth]{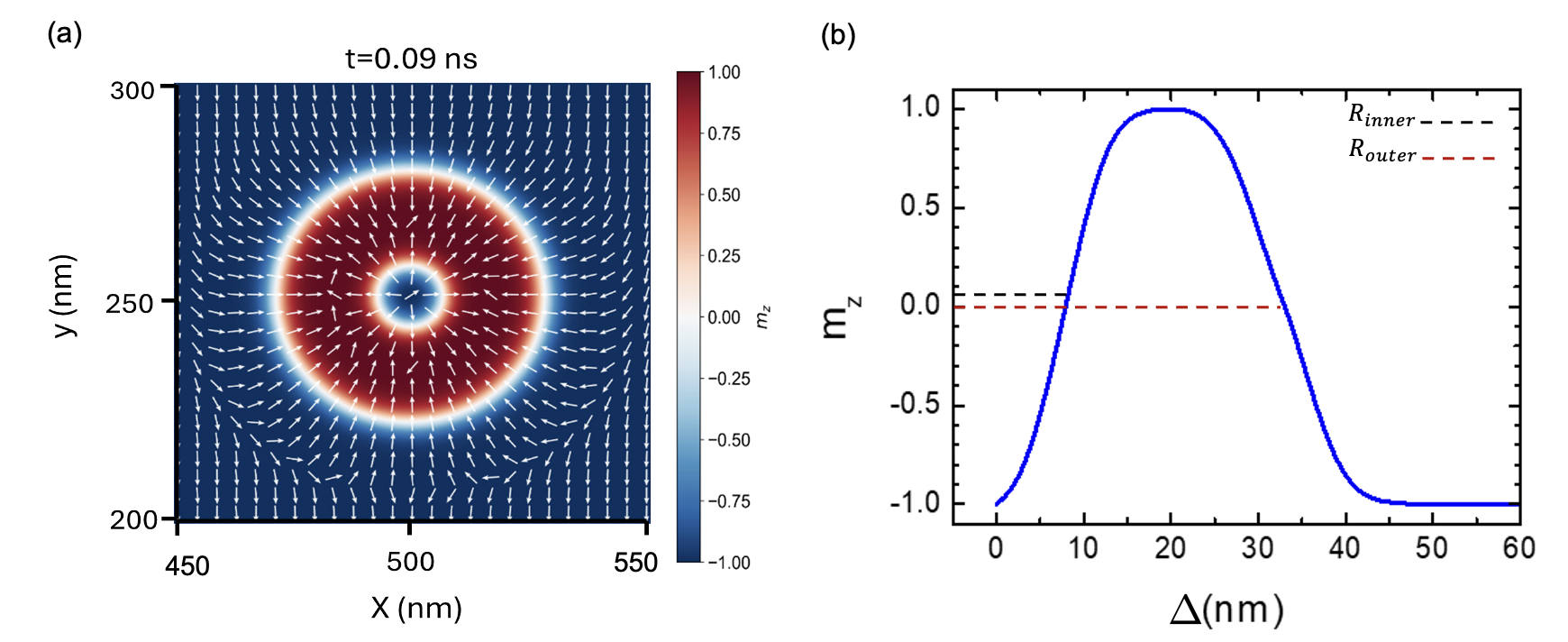}
\caption{
Formation of a skyrmionium structure after the merging of two N\'eel skyrmions. 
(a) Magnetization configuration showing the characteristic ring-like structure with a reversed core. 
(b) Radial profile $m_z(r)$ exhibiting two zero crossings, which define the inner radius $R_{\mathrm{inner}}$ and outer radius $R_{\mathrm{outer}}$ of the skyrmionium.
}
\end{figure}

At short separations ($\Delta \le R$) as in Fig~2 (d), the two skyrmions strongly overlap and their internal structures merge. 
In this limit, the outcome depends sensitively on the relative winding number and polarity. We observed that, for all combinations except the case of the same polarity and opposite winding number, the system relaxes to a single skyrmion as shown in Fig~3 (a). Panel (b) shows the radial profile of the out-of-plane magnetization component $m_z(r)$ extracted through the center of the merged structure. The profile exhibits a smooth transition from $m_z \approx +1$ at the core to $m_z \approx -1$ in the ferromagnetic background, indicating the formation of a single skyrmion after the merging process. The skyrmion radius $R$ is defined as the distance from the center to the
domain wall, identified by the condition $m_z=0$. The profile crosses
$m_z=0$ at a distance of approximately $15~\mathrm{nm}$ from the center, yielding a skyrmion radius of $R\simeq15~\mathrm{nm}$.

Compared to the initial skyrmions each with radius $R = 23$ nm, the reduced width of the $m_z(r)$ profile demonstrates that the merged configuration is more compact. This reduction in size reflects the relaxation of the system toward a lower-energy state, where the exchange and anisotropy energies favor a narrower domain wall and a smaller skyrmion. The absence of additional zero crossings in $m_z(r)$ further confirms that the final state corresponds to a single skyrmion rather than a skyrmionium, which would exhibit a double-domain-wall structure.

For Néel skyrmions with the same polarity but opposite winding numbers (topological indices \(n\)), a qualitatively different merging behavior is observed, as illustrated in Fig. 4(a). In a Néel-type magnetic texture, the magnetization rotates predominantly within the local radial–out-of-plane (\(r\)–\(z\)) plane across the domain wall. When the two skyrmions partially overlap, their magnetization distributions undergo a substantial reconfiguration in the interaction region, leading, under suitable conditions, to the formation of a concentric double-domain-wall structure characteristic of a skyrmionium.

Unlike a conventional skyrmion, which exhibits a radial magnetization rotation of approximately \(\pi\), the resulting skyrmionium accommodates a complete \(2\pi\) rotation of the magnetization from its center to the surrounding ferromagnetic background. This configuration consists of two concentric domain walls and is characterized by a vanishing net topological charge, \(Q=0\) \cite{Zhang2015,Zhang2018,Zhang2016,Gobel2021}.

\begin{figure}[b!]
    \centering
    \includegraphics[width=1\linewidth]{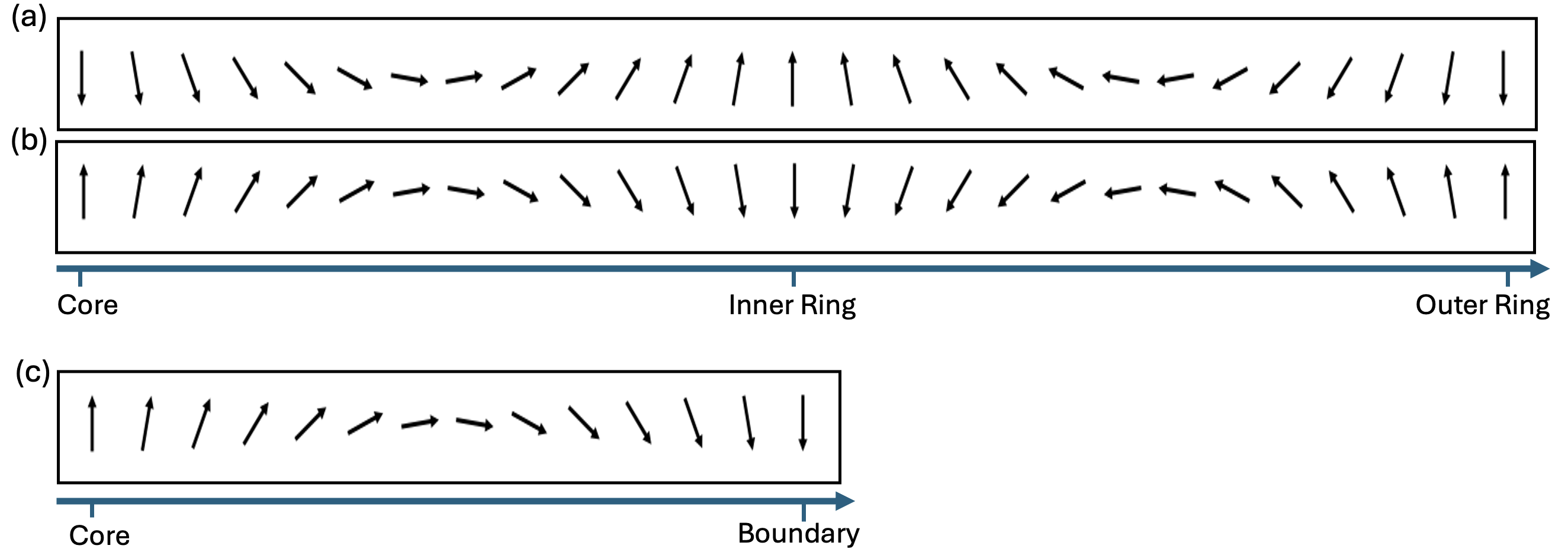}
\caption{Schematic cross-sectional representation of spin orientations in (a, b) skyrmionium configurations with opposite core polarities and (c) an isolated skyrmion.
}
\end{figure}

An examination of the resulting magnetic configurations reveals distinct spin-reorientation profiles associated with skyrmionium formation. As illustrated in Figs 5 (a) and (b), the skyrmionium exhibits a double-domain-wall structure, in which the out-of-plane magnetization reverses twice along a cross-sectional path through its center. For a downward-oriented ferromagnetic background, the magnetization follows the sequence \(m_z=-1\rightarrow+1\rightarrow-1\), whereas reversing the background magnetization and core polarity produces the opposite sequence, \(m_z=+1\rightarrow-1\rightarrow+1\). In contrast, the isolated skyrmion shown in Fig 5 (c) exhibits a single magnetization reversal between its core and the surrounding background.

For interacting skyrmions with opposite winding numbers, the merger process can reconstruct the initial spin textures into a skyrmionium configuration, with the inner and outer rings exhibiting opposite winding numbers. Importantly, the assignment of these winding numbers to the inner and outer rings plays a crucial role in determining the final magnetic configuration. In the present case, the inner ring exhibits \(w_{\mathrm{in}}=+1\), while the outer ring exhibits \(w_{\mathrm{out}}=-1\). This arrangement allows the outer region to remain compatible with the downward-oriented ferromagnetic background, supporting the formation of a skyrmionium.

However, when the initial positions of the two interacting skyrmions are exchanged, the winding-number assignments of the inner and outer rings are reversed. Consequently, the inner ring becomes compatible with the surrounding ferromagnetic background rather than the outer ring. This compatibility allows the inner region to merge with the background magnetization, leading to the disappearance of the inner domain wall and the transformation of the configuration into a single skyrmion.

These observations suggest that skyrmionium formation depends not only on the opposite winding numbers of the interacting skyrmions but also on their initial spatial arrangement and the resulting distribution of winding numbers between the inner and outer rings. The results further highlight the importance of the initial winding-number configuration in determining the final magnetic state \cite{Gobel2021,lin2013,Nagaosa2013}.
Since skyrmionium formation is observed for N\'eel skyrmions with identical polarity and opposite winding number, the subsequent analysis focuses on this specific configuration. We observed that it is the only one that leads to the formation of a stable skyrmionium state. The results presented below, including the time evolution of the energy contributions as well as the skyrmion displacement and velocity, are therefore analyzed within this specific configuration in order to clarify the physical mechanism underlying skyrmionium formation. Figure~4 (b) shows the radial profile of the out-of-plane magnetization component $m_z(r)$ taken through the center of the structure. Unlike a single skyrmion, which exhibits a single zero crossing, the profile here displays two distinct zero crossings, indicating the presence of a skyrmionium (a $2\pi$ topological structure). 

Starting from the center ($\Delta = 0$), the magnetization is maximally negative ($m_z = -1$), forming a reversed core. As $\Delta$ increases, $m_z(r)$ rises and crosses zero at the first point, defining the inner radius $R_{\mathrm{inner}}$. This corresponds to the first domain wall separating the inner core from the surrounding ring where the magnetization reaches the positive maximum ($m_z = +1$). 

Further away from the center, $m_z(r)$ decreases again and crosses zero at a second point, defining the outer radius $R_{\mathrm{outer}}$, which corresponds to the second domain wall. Beyond this point, the magnetization returns to the uniform background state ($m_z = -1$). 

This double-domain-wall structure is the hallmark of a skyrmionium, where the spin texture undergoes a full $2\pi$ rotation. The presence of two zero crossings in $m_z(r)$ clearly distinguishes this state from a single skyrmion and confirms the formation of a stable skyrmionium configuration.
 The inner radius is observed to be $R_{\mathrm{in}} = 10$ nm, while the outer radius one is $R_{\mathrm{out}} = 34$ nm with the latter defining the overall skyrmionium size. Compared with the initial merging skyrmions of radius $23$ nm, the outer radius of the final skyrmionium is larger by about $11$ nm, indicating that the merging process leads to an expanded ring-like topological structure rather than a compact single skyrmion.

 The formation of the skyrmionium should not be interpreted as a simple
geometrical overlap of two skyrmions. Once the domain walls interact, the
initial textures lose their individual identities and the complete
magnetization field undergoes a nonlinear reconstruction. The final state is
therefore determined by whether the reduction of exchange frustration and
magnetostatic energy can compensate for the energetic cost of creating an
additional domain wall and an extended region of tilted magnetization.

\begin{figure}[bt!]
\centering \includegraphics[width=1\linewidth]{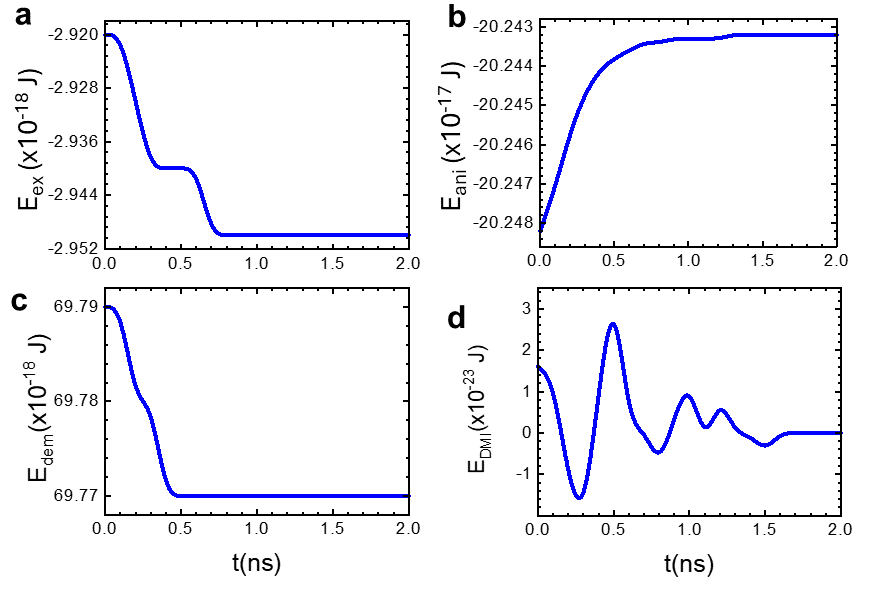}
\caption{Time evolution of the energy contributions during the relaxation process of skyrmenioum: (a) exchange energy, (b) anisotropy energy, (c) demagnetizing energy, and (d) DMI energy. The exchange and demagnetizing energies decrease with time, indicating a reduction in magnetization gradients and stray-field energy as the spin texture becomes smoother. In contrast, the anisotropy energy increases and approaches a steady value, reflecting the alignment of spins along the easy axis. All energy terms converge toward stable plateaus, demonstrating the relaxation of the system to an energetically favorable equilibrium state.}\label{fig:FIG5.png} \end{figure}

Figure~6 reveals the microscopic mechanism governing the
transformation of two partially overlapping N\'eel skyrmions into a
skyrmionium. The individual energy contributions do not evolve independently;
rather, they exchange energy cooperatively as the initially frustrated overlap
region reconstructs into a concentric double-domain-wall texture.

At the beginning of the relaxation, the two skyrmion profiles occupy the same
spatial region while attempting to preserve their individual radial spin
rotations. This produces strong spatial gradients and a locally incompatible
magnetization distribution within the overlap region. Because the exchange
energy density scales as $A\sum_i\left(\nabla m_i\right)^2,$
these large gradients generate a substantial exchange-energy penalty. The
rapid decrease of $E_{\mathrm{ex}}$ observed in
Fig.~6(a) therefore identifies the removal of exchange
frustration as the principal driving mechanism of the initial merging stage.
As the magnetization relaxes, the highly compressed and distorted spin
gradients are redistributed over a broader annular region, allowing the two
initial textures to reconstruct into an inner and an outer domain wall.
The first decrease in $E_{\mathrm{ex}}$ corresponds predominantly to the
elimination of the most strongly frustrated spins in the central overlap
region, whereas the subsequent smaller decrease reflects the slower
reorganization of the domain-wall profiles toward their equilibrium widths
and radii. The final plateau indicates that the magnetization gradients have
reached a stationary distribution compatible with a metastable
skyrmionium.

The evolution of the anisotropy energy,
Fig.~6(b), provides complementary information about the
geometrical reconstruction. For perpendicular anisotropy, the local energy
density is proportional to$K_{\mathrm{eff}}\left(1-m_z^2\right),$
and is therefore largest in regions where the magnetization tilts away from
the easy axis. During skyrmionium formation, the initially localized
transition regions expand into two concentric domain walls. Consequently,
the total area occupied by strongly tilted spins increases, producing the
observed increase in anisotropy energy. This increase should not be
interpreted as evidence of an unstable final state. Instead, it represents
the energetic cost required to create and maintain the extended
double-domain-wall structure. The system accepts this anisotropy penalty
because it is more than compensated by the larger reduction in exchange and
magnetostatic energies.

The decrease of the demagnetizing energy shown in
Fig.~6(c) reflects the progressive redistribution of magnetic
charges during the merging process. In the initial partially overlapped
configuration, the distorted domain walls and asymmetric magnetization
distribution generate nonuniform surface and volume magnetic charges,
producing a relatively large stray field. As the skyrmionium forms, the
magnetization becomes more circularly symmetric and the magnetic charges are
distributed more uniformly over the concentric walls. This improved flux
closure lowers the magnetostatic energy and contributes to the stabilization
of the final ring-like configuration. Although the magnitude of this change
is smaller than the exchange-energy reduction, it provides an important
secondary stabilization mechanism.

In contrast to the largely monotonic exchange, anisotropy, and demagnetizing
contributions, the DMI energy in Fig.~6(d) exhibits damped
oscillations . This behaviour originates from repeated local readjustments of
the chirality during the reconstruction of the two initial N\'eel textures.
The interfacial DMI favours a specific rotational sense of the magnetization
across each domain wall. During the early merging stage, however, the
overlapping spin textures are not everywhere compatible with this preferred
chiral arrangement. The domain walls therefore undergo successive
compression, expansion, and azimuthal reorientation as the magnetization
searches for a configuration that simultaneously reduces exchange gradients
and preserves the preferred N\'eel chirality. The oscillatory DMI response is
thus the energetic signature of this transient chiral reconstruction. Its
progressive damping indicates that the local rotational sense becomes
stabilized as the inner and outer walls approach their equilibrium
configuration.

Examining the Fig.6 systematically, we arrive at a conclusion that the energy evolution can be separated into two characteristic
stages. The first is an exchange-dominated reconstruction stage, during which
the system rapidly removes the intense magnetization gradients created by
the initial overlap. The second is a slower profile-relaxation stage, during
which the radii, widths, and chiral structure of the two domain walls adjust
toward their equilibrium values. The simultaneous saturation of all energy
contributions demonstrates that the final skyrmionium is not a frozen
geometrical superposition of two skyrmions, but a dynamically reconstructed
metastable state corresponding to a local minimum of the complete
micromagnetic energy functional.

The initial size of the interacting skyrmions plays a decisive role in
determining the final magnetic configuration after merging. In particular,
the formation of a stable skyrmionium is not governed solely by the polarity
and winding-number combination, but also requires the interacting skyrmions
to retain the characteristic radius supported by the chosen material
parameters. For the favorable configuration of identical polarity and
opposite winding number, a stable skyrmionium is obtained when both
skyrmions have this characteristic radius. Thus, skyrmionium formation
should not be interpreted simply as a consequence of having sufficiently
large skyrmions; rather, it occurs within a specific size condition determined
by the material parameters.\\
A notable change in the merging pathway occurs when the radius of only one
of the interacting skyrmions is reduced or increased, even slightly, while the other
parameters are kept unchanged. Although the polarity and winding-number
combination remains favorable for skyrmionium formation, the size mismatch
prevents the overlapping textures from reconstructing into a stable
concentric double-domain-wall configuration. Instead, the system relaxes
into a single-skyrmion state. This demonstrates that the topological
selection rule alone is not sufficient to determine the final state; the
relative sizes of the interacting skyrmions provide an additional
geometrical constraint on the merging pathway.\\
Upon further reduction of the skyrmion radius, a second transition is
observed. For the present material parameters, when the reduced radius
reaches approximately $18~\mathrm{nm}$, the merged texture no longer
survives and undergoes annihilation, as shown in Fig.~6. The time-resolved
magnetization snapshots in Fig.~7(a) show a progressive contraction of the
magnetic texture followed by the disappearance of the skyrmion core and
relaxation toward the nearly uniform ferromagnetic background. This behavior
is also reflected in Fig.~7(b), where the effective radius continuously
decreases and eventually approaches zero.

These results reveal a size-dependent sequence of merging outcomes under
otherwise identical interaction conditions. When both skyrmions retain the
material-supported radius, the identical-polarity/opposite-winding
configuration can reconstruct into a stable skyrmionium. A reduction in the
radius of one of the interacting skyrmions redirects the merging pathway
toward a single-skyrmion state, while a further reduction to approximately
$18~\mathrm{nm}$ leads to complete annihilation. The final state is therefore
determined by the combined influence of topology, initial separation, and
the size of the interacting skyrmions.
\begin{figure}[tb!]
\centering \includegraphics[width=1\linewidth]{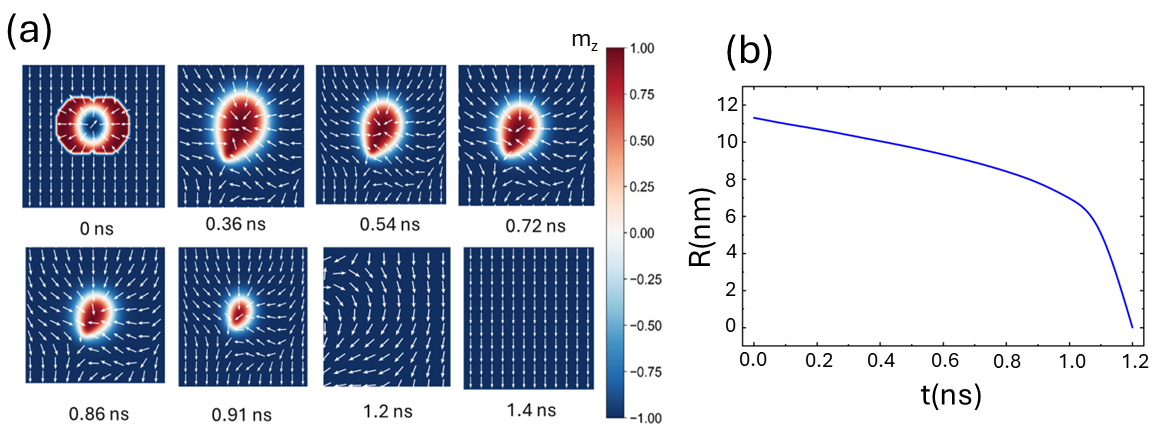}
\caption{
Time evolution of two skyrmions with identical polarity and opposite winding number, initially separated by a distance $\Delta=R$, illustrating their merging dynamics in the reduced-size regime. 
}\label{fig:FIG1.png} \end{figure}

To further confirm the stability of the skyrmionium state formed after merging, we analyze its time-dependent dynamics in the low-damping regime. Figure~7 shows the temporal evolution of the magnetization configuration together with the extracted inner radius $R_{\mathrm{inner}}$, outer radius $R_{\mathrm{outer}}$, and effective radius $R_{\mathrm{eff}}$. The skyrmionium preserves its circular symmetry while undergoing periodic expansion and contraction, characteristic of a breathing mode, as clearly observed in Fig.~8(a), where the magnetization configuration remains topologically intact with no evidence of collapse or transformation into a single skyrmion. The radial analysis in Fig.~8(b) demonstrates that both $R_{\mathrm{inner}}$ and $R_{\mathrm{outer}}$ oscillate around well-defined mean values, indicating that the double-domain-wall structure is maintained, while the effective radius $R_{\mathrm{eff}}$ exhibits a clear sinusoidal behavior, as shown in Fig.~8(c), confirming the presence of a coherent breathing mode. Importantly, the oscillations remain bounded and do not grow with time, indicating that the skyrmionium corresponds to a dynamically stable state rather than a transient configuration. This behavior reflects the balance between exchange, anisotropy, Dzyaloshinskii--Moriya interaction, and dipolar energies, which collectively stabilize the $2\pi$ spin structure. The bounded oscillatory behavior is also consistent with the energy relaxation observed in Fig.~6, where all energy contributions converge to stable plateaus, indicating that the system has reached a local energy minimum supporting a stable skyrmionium configuration. To quantitatively characterize the breathing dynamics, we analyze the time evolution of the effective radius $R(t)$, which follows a harmonic form $R(t) = R_0 + \delta R \sin(\omega t)$. From the numerical simulation, we extract $R_0 = 21.463~\mathrm{nm}$ and $\delta R = 0.74~\mathrm{nm}$, indicating that the oscillations occur around a well-defined equilibrium radius with a relatively small amplitude, while the corresponding breathing frequency is $f = 11.76~\mathrm{GHz}$. The condition $\delta R \ll R_0$ confirms that the skyrmionium undergoes a weakly nonlinear oscillation around its equilibrium configuration, reflecting the restoring forces arising from the competition between exchange, anisotropy, Dzyaloshinskii--Moriya interaction, and dipolar energies. The observed GHz frequency is characteristic of intrinsic eigenmodes of confined magnetic textures, and its persistence further supports that the skyrmionium corresponds to a dynamically stable state, with the breathing mode representing a natural excitation of the double-domain-wall structure. These results demonstrate that the skyrmionium formed via merging is not only energetically favorable but also dynamically robust, and provide further evidence that its formation for the case of identical polarity and opposite winding number corresponds to a true equilibrium state consistent with the topological selection rule identified in this work.

\begin{figure}[tb!]
\centering \includegraphics[width=0.8\linewidth]{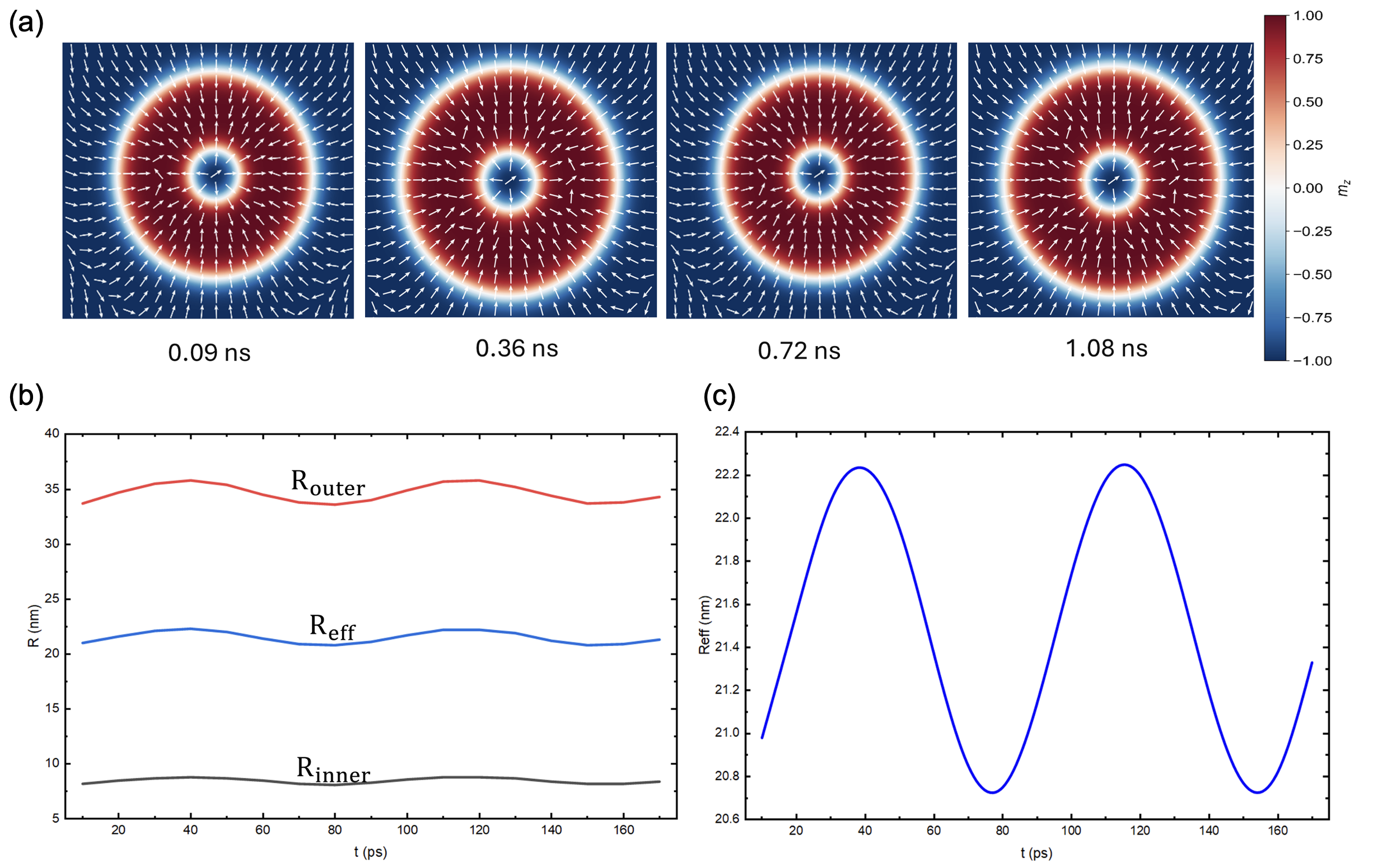}
\caption{Time evolution of a skyrmionium exhibiting a radial breathing mode.  
As time progresses, the skyrmionium preserves its circular symmetry while undergoing periodic expansion and contraction of both its inner and outer radii. 
This behavior corresponds to a breathing mode, where the two concentric domain walls oscillate coherently without translational motion of the texture center.}
\label{fig:FIG1.png} \end{figure}

\section{Conclusion}
In this work, we have established the key physical conditions governing
the nonlinear merging of two N\'eel skyrmions and the emergence of a
stable skyrmionium. The initial separation plays a decisive role in
setting the interaction regime: widely separated skyrmions interact
weakly, whereas decreasing the separation enhances the overlap of their
spin textures and drives the system into a strongly nonlinear regime
where merging and topological reconstruction become possible.

Within this regime, the final state is selected by the internal
topological configuration of the interacting skyrmions. A stable
skyrmionium emerges for skyrmions with identical polarity and opposite
winding numbers, while the other investigated configurations relax
predominantly into a single-skyrmion state. The initial skyrmion size
provides an additional geometrical constraint, demonstrating that the
topologically favorable configuration alone is insufficient when the
textures are too small to sustain the resulting double-domain-wall
structure.

The energy-resolved dynamics show that skyrmionium formation is a
nonlinear reconstruction process driven by the cooperative
redistribution of exchange, anisotropy, Dzyaloshinskii--Moriya, and
magnetostatic energies. The resulting skyrmionium remains dynamically
stable and exhibits a bounded radial breathing mode about its
equilibrium size. These results demonstrate that merger-induced
skyrmionium formation is collectively controlled by initial separation,
topological compatibility, skyrmion size, and energetic relaxation,
providing a predictive framework for engineering higher-order magnetic
textures through controlled skyrmion interactions.
\section*{Acknowledgments}
R. S. would like to acknowledge the support from the HMTF Strategic Research of Oman
(Grant No. SR/—/SCI/P/26/648). The authors acknowledge New York University Abu Dhabi (NYUAD) for providing computational resources used in carrying out the micromagnetic simulations presented in this work.
\vspace{-0.5em}

\FloatBarrier
\bibliographystyle{apsrev4-2}
\bibliography{References}

\end{document}